# Robust Mottness and tunable interlayer magnetism in $Nb_3X_8$ (X = F, Cl, Br, I) bilayers

Zhongqin Zhang[1,2], Jiaqi Dai[1,2], Cong Wang[1,2], Zhihai Cheng[1,2], and Wei Ji[1,2*]

[1]*Beijing Key Laboratory of Optoelectronic Functional Materials & Micro-Nano Devices, School of Physics, Renmin University of China, Beijing 100872, China*

[2]*Key Laboratory of Quantum State Construction and Manipulation (Ministry of Education), Renmin University of China, Beijing 100872, China*

* W.J. (email: wji@ruc.edu.cn)

**Abstract:**

Kagome materials have attracted extensive attention due to their correlated properties. The breathing kagome material system $Nb_3X_8$ (X = F, Cl, Br, I) is regarded as a Mott insulator. However, studies on the influence of interlayer coupling on its magnetic and Mott properties are lacking. In this work, we investigated the effect of interlayer coupling on bilayer properties of each $Nb_3X_8$ (X = F, Cl, Br, I) compound via density functional theory (DFT) calculations, considering 24 stacking configurations per material. We found that each bilayer material is a Mott insulator. Due to the competition between interlayer Pauli repulsion and hopping, most interlayer magnetism is AFM, a small number of cases show AFM-FM degeneracy, and the magnetic ground state of 3 configurations is interlayer FM, i.e., tunable interlayer magnetism occurs. This robustness of Mott states coexisting with tunable interlayer magnetism provide novel and comprehensive analysis and insights for the research of breathing kagome Mott insulators.

Keywords: kagome lattice; Mott insulator; van der Waals magnets; $Nb_3Cl_8$; $Nb_3F_8$; $Nb_3Br_8$; $Nb_3I_8$.

Kagome-lattice materials are exceptional material platforms for exploring the interplay among electronic band topology, strong electron correlation, and magnetism in quantum materials[1–7]. The standard kagome tight-binding model is in two-dimensional (2D) and exhibits a flat band across the entire Brillouin zone, Dirac cones at the K points, and van Hove singularities located at the M points. Theoretical studies predict that tuning electron filling within these kagome bands can stabilize diverse exotic phases, such as magnetism[8,9], superconductivity[10,11], charge-density wave states[12]. Experimental investigations, however, have primarily focused on three-dimensional (3D) bulk kagome materials, such as $CsV_3Sb_5$[7,13,14] and $Co_3SnS_2$ [15–17], where superconductivity[13,14], pair-density wave[7], charge order[13], and spin-orbit polarons[15] have been reported. In bulk crystals, intrinsic kagome layers typical alternate with spacer layers, whose strong interlayer hybridization[18] often shifts or even eliminates the characteristic kagome bands near the Fermi level[18,19], obscuring the exploration of intrinsic kagome physics.

Significant effort has recently been devoted to constructing 2D kagome lattices with minimal interlayer coupling [4]. Strategies include surface-assembled organic kagome monolayers[20–22], $MoTe_{2-x}$ monolayers mirror-twin-boundary loops [23–25], atomic intercalation within van der Waals (vdW) layers[26,27], and a recently highlighted 3+1 protocol [28]. Among them, $Nb_3X_8$ (X = Cl, Br, I) is a prototypical layered breathing kagome family, with adjacent layers weakly coupled through vdW interactions[29–35]. Monolayers $Nb_3Cl_8$ and $Nb_3Br_8$ are established Mott insulators, characterized by well-defined upper and lower Mott-Hubbard bands around the Fermi level[31,36]. Their bulk counterparts, however, exhibit more complicated behaviors, particularly in magnetism[34,35]. Bulk $Nb_3Cl_8$ and $Nb_3Br_8$ undergo structural transitions at approximately 100 K and 380 K, respectively, accompanying by a magnetic transition from Curie-Weiss paramagnetism (PM) in the high-temperature (HT) phase to a likely nonmagnetic (NM) state at lower temperatures (LT) [34,35]. Interestingly, these contrasting magnetic states differ structurally in their stacking sequences only. Moreover, the two previously proposed mechanisms for the LT NM state are both closely related to interlayer couplings[34,35], further highlighting the importance of interlayer interactions in $Nb_3X_8$.

Interlayer coupling, a unique feature of 2D materials, has increasingly recognized as an effective route to tune electronic band gaps[37,38], optical[39,40] and acoustic properties[41], magnetism[42–45], and electric polarization[46,47]. Recently, it was found that interlayer coupling in bilayer 1T-$NbSe_2$ can transform it from a correlated insulator into a nonmagnetic band insulator, raising a crucial question: what is the role of interlayer coupling on electronic structures of kagome-based $Nb_3X_8$ bilayers? In this letter, we theoretically investigated the roles of interlayer coupling in tuning the electronic structures and magnetism in $Nb_3X_8$ (X = F, Cl, Br, I) bilayers using density functional theory (DFT) calculations. We first identified out-of-plane (OOP) electric polarization and a robust Mott-Hubbard gap in $Nb_3X_8$ monolayers using spin polarized bandstructure calculations with varying $U$. Next, we compared the total energies of 24 stacking configurations for each compound among NM and interlayer AFM and FM states. Most bilayers favor interlayer AFM, while a few exhibit AFM-FM degeneracy. All bilayers remain Mott insulators, as verified by real-space wavefunction distributions, bandstructures, and $U$ dependence. Symmetry analyses uncover compensated AFM states in parallel electric polarized stackings, classifying all studied configurations into three stacking-dependent and tunable interlayer magnetism. This robustness of Mott states coexisting with tunable interlayer magnetism contrasts sharply with the behaviors of transition-metal dichalcogenide-based Mott insulators.

Our density functional theory (DFT) calculations were performed using the generalized gradient approximation (GGA) for the exchange correlation potential in the form of Perdew-Burke-Ernzerhof (PBE)[48], the projector augmented wave method[49], and a plane-wave basis set as implemented in the Vienna ab-initio simulation package (VASP)[50]. The dispersion correction was included using Grimme's semiempirical D3 scheme[51] in combination with the PBE functional (PBE-D3). A kinetic energy cutoff of 450 eV for the planewave basis sets was adopted for all structural relaxations and electronic structure calculations. All atoms were allowed to relax until the residual force per atom is below 0.01 eV/Å. A vacuum layer exceeding 15 Å in thickness was used to reduce imaging interactions from neighboring supercells. Gamma-centered Monkhorst-Pack $k$-meshes of 5×5×1 and 9×9×1 were used to sample the Brillouin zone of 1×1 unit cells for structural relaxations and electronic structure calculations, respectively. In comparison the energies of magnetic configurations, a $2\times\sqrt{3}$ supercell and a 4×5 $k$-mesh were employed. A Gaussian smearing width of 0.01 eV was use to smooth the distribution function of electronic states in all calculations. A DFT+$U$ method implemented using the rotationally invariant approach[52] was used to apply a series of Hubbard $U$ values specifically to Nb 3$d$ orbitals. Comparison of energies of different magnetic configurations were made based on the structures optimized for interlayer AFM coupling. Energy comparison based on the structures relaxed using interlayer FM does not change the relative energetic stability.

A free-standing $Nb_3X_8$ monolayer adopts a breathing kagome geometry [Fig. 1(a) and 1(b)], in which compact $Nb_3$ trimers (shaded in orange) corner-sharing with expanded Nb triangles (shaded in gray), giving rise to alternating short ($d_{Nb1}$) and long ($d_{Nb2}$) Nb–Nb bonds. The halogen atoms bonded to these triangular units form two inequivalent sublayers. In the lower sublayer, three equal-height halogen atoms lie beneath the expanded Nb triangle, giving a shorter vertical distance to the Nb plane ($d_{XS}$) than that of the halogens in the upper sublayer ($d_{XL}$). The remaining halogen atom in the lower sublayer, residing beneath the compact trimer [X-s, purple, Fig. 1(a) to 1(c)], is displaced further away from the Nb plane by $\varDelta_S$ relative to the other three halogens in the same sublayer. In the upper sublayer, the halogen sitting above the expanded triangle (X-l, pink) is shifted closer to the Nb plane by $\varDelta_L$ compared with the other three upper-sublayer halogens. We denote the lower and upper sublayers as X-s and X-l, respectively.

For $Nb_3F_8$, the Nb-Nb bond length in the $Nb_3$ trimer ($d_{Nb1}$) contracts to 2.59 Å, well below the Nb-Nb bonds in bulk Nb (approximately 2.86 Å), whereas in the other

Nb halides, $d_{Nb1}$ is comparable to or larger than the bulk vale. This pronounced contraction of $d_{Nb1}$ indicates strong trimer compression and thereby enhances the X-s displacement ($\Delta_S$). Table 1 compiles the two ratios $\Delta_S/d_{XS}$ and $\Delta_L/d_{XL}$ for all four Nb halides. With the sole exception of $\Delta_S/d_{XS}$ in $Nb_3F_8$ (0.44), which is anomalously large, the other seven entries decrease monotonically as X is varied from I to F. These opposite out-of-plane (OOP) displacements ($\Delta_S$ and $\Delta_L$) and inequivalent distances ($d_{XS}$ and $d_{XL}$) lead to a net OOP electric polarization for each Nb halide, with $Nb_3F_8$ exhibiting the largest value of 1.26 pC/m (Table 1, Fig. 1(c)).

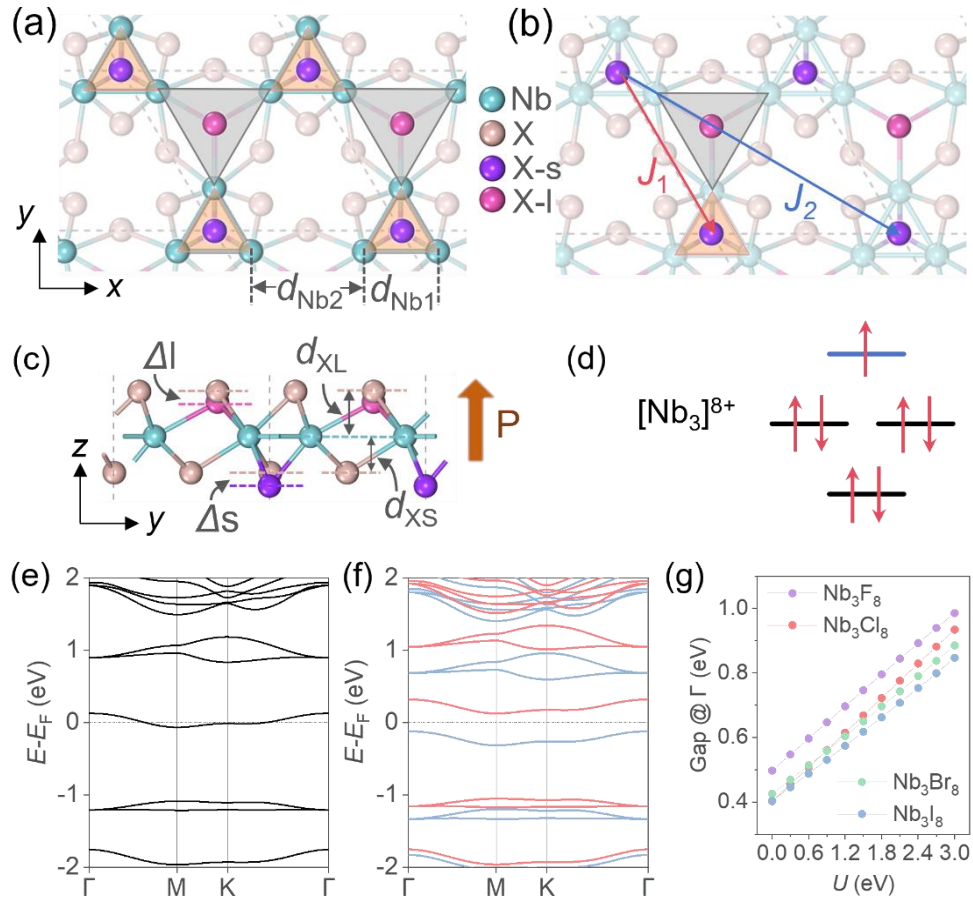

Fig. 1. (a-c) Top(a-b) and side(c) view of monolayer $Nb_3X_8$. Nb atoms are represented by light sea green balls. The gray and orange shading indicate the large and small triangles formed by Nb atoms, with X atoms at the center of these triangles labeled as X-l and X-s, shown by pink and purple balls, and remaining X atoms are represented by light brown balls. The brown arrow and "P" denote the polarization direction. Red and blue arrows represent the nearest-neighbor and next-nearest-neighbor exchange couplings between $Nb_3$ trimers. Gray dash lines indicate the Nb-Nb distance within ($d_{Nb1}$) and between ($d_{Nb2}$) trimers. $d_{XS}$ and $d_{XL}$ represent the vertical displacements of corresponding X relative to Nb atomic layer, while $\Delta_S$,

$\varDelta_L$ represent the vertical displacements of X-s and X-l atoms relative to the other three halogens in the same sublayer. (d) Electronic configuration of $[Nb_3]^{8+}$. (e-f) Band structures of $Nb_3Cl_8$ in NM (e) and FM (f) configurations. In (f), the blue and red bands represent the spin-up and spin-down channels. (g) Band gaps at the Γ point as a function of Hubbard $U$, calculated using the magnetic ground states.

We compared the total energies of seven magnetic configurations for each $Nb_3X_8$ monolayer for finding the most favored magnetic configuration among them, as shown in Supplementary Fig. S1. Triangularly arranged Nb atoms in a $Nb_3$ trimer have geometric frustration, which, however, favor a FM nearest neighboring coupling, thereby exhibiting a shared local magnetic moment on the trimer. Therefore, we treated the $Nb_3$ trimer as a single entity with the +8 oxidation state. The valence electron configuration of the trimer, as shown in Fig. 1d, includes one unpaired electron and six paired electrons, exhibiting a magnetic moment of 1 $\mu_B$. Thus, the triangularly arranged trimers form a $S = 1/2$ triangular lattice site. While either the $Nb_3Br_8$ or $Nb_3I_8$ monolayer favors an FM configuration, $Nb_3Cl_8$ has nearly degenerate FM, ZZ-AFM and Stripe-AFM configurations, consistent with the literature [31,53,54]. However, the ZZ-AFM configuration is slightly stabilized in $Nb_3F_8$ (0.25 meV/$Nb_3$ more stable than the stripe-AFM configuration). Based on energy comparisons among the various magnetic configurations, we calculated the nearest-neighbor ($J_1$) and next-nearest-neighbor ($J_2$) spin-exchange coupling constants among these $Nb_3$ trimers[indicated by red and blue arrows in Fig. 1(b)], which are summarized in Table 1. Each $Nb_3X_8$ monolayer exhibits easy-plane anisotropy, with the magnetic anisotropy energy ranging from 0.11 to 0.38 meV/$Nb_3$ [Fig. S2, Table S2].

Monolayer $Nb_3Cl_8$ [31–33] or $Nb_3Br_8$ [29,30], has been identified as a Mott insulator in previous dynamical mean field theory (DMFT)[31,32] and model-based [29,30,33] calculations. A spin non-polarized DFT calculation for, for instance, $Nb_3Cl_8$ reveals a partial (half)-occupied spin-degenerate quasi-flat band that cuts the Fermi level, as plotted in Fig. 1(e). This degeneracy lifts in a spin polarization bandstructure with the FM configuration, splitting the partial occupied band into a fully occupied and an empty spin-polarized band [Fig. 1(f)], as a result of the electron exchange effect. To capture strong electron correlations, incorporating an additional on-site Hubbard $U$ energy corrects the self-interaction error of DFT and localizes electronic wavefunctions, and further enhances the exchange splitting. Consequently, the split gap at the Γ point grows linearly as a function of $U$ [Fig. 1(g)], exhibiting a signature characteristic of Mott insulators. Our DFT+U results thus confirm that all four $Nb_3X_8$ monolayers are Mott insulators, consistent with previous DMPT results [31–33,55,56].

Table 1 Lattice constants ($a$), Nb-Nb distance within ($d_{Nb1}$) and between ($d_{Nb2}$) $Nb_3$ trimers, vertical shift of X atoms (see Fig. 1(c)), polarization ($P$, in units of pC/m), magnetic ground states (G. S.) and the nearest-neighbor ($J_1$) and next-nearest-neighbor ($J_2$) exchange couplings between Nb3 trimers (in units of meV/$Nb_3$) for monolayer $Nb_3X_8$. Here, the magnetic coupling between Nb atoms within a $Nb_3$ trimer is FM. A positive value of the exchange constant indicates AFM coupling.

| Halides | $a$ (Å) | $d_{Nb1}$ (Å) | $d_{Nb2}$ (Å) | $\Delta_S$ (Å) | $d_{XL}$ (Å) | $\Delta_S/d_{XL}$ | $\Delta_S$ (Å) | $d_{XL}$ (Å) | $\Delta_S/d_{XL}$ | $P$ | G. S. | $J_1$ | $J_2$ |
|---|---|---|---|---|---|---|---|---|---|---|---|---|---|
| $Nb_3F_8$ | 5.95 | 2.59 | 3.36 | 0.48 | 1.10 | 0.44 | 0.23 | 1.35 | 0.17 | 1.26 | ZZ | 6.75 | -0.49 |
| $Nb_3Cl_8$ | 6.76 | 2.81 | 3.95 | 0.47 | 1.36 | 0.35 | 0.29 | 1.64 | 0.18 | 0.21 | FM | 0.12 | -0.24 |
| $Nb_3Br_8$ | 7.11 | 2.88 | 4.23 | 0.54 | 1.44 | 0.38 | 0.35 | 1.76 | 0.20 | 0.41 | FM | -2.74 | -0.37 |
| $Nb_3I_8$ | 7.63 | 3.01 | 4.62 | 0.60 | 1.54 | 0.39 | 0.47 | 1.91 | 0.25 | 0.56 | FM | -2.93 | -0.82 |

We next investigate how interlayer couplings reshape the geometric and electronic structures of bilayer $Nb_3X_8$. As the monolayer exhibits an intrinsic OOP polarization, the two layers can stack in three alignments, namely up-up (Fig. S9(h)), up-down [Fig. 2(a)], and down-up (Fig. S4(b-c)), which are label as UU, UD, and DU, respectively. Lateral stacking registries further differentiates the stackings. Figures 2(b) and 2(c) present the top-views of the up- and down-polarized bottom $Nb_3X_8$ layers, respectively, in which interfacial X and Nb atoms are highlighted and the outer X atoms are faded for clarity. We identify four inequivalent interfacial X triangles (shadowed in orange) with their hollow sites labelled 1 to 4, in either up- or down-polarization case. Taking the X-s atom of the top layer as a reference for alignment UD (for alignments UU and

DU, we take the X-l atom), sliding the top layer to align the X-s (or X-l) atom [Fig. 2(a)] occupying at any one of these sites constructs four distinct stacking configurations. A third degree of freedom is the relative twist between the two layers. We considered untwisted (R0) and 60° twisted (R60 bilayers, as shown in Fig. 2(d) and 2(e), respectively. Combining polarization (three alignments) with hollow site (four sites) and twist angle (two angles) yields 24 distinct stacking configurations per halide. We label them as (site)-(twist)-(polarization), for instance, 1-R0-UU corresponds to the AA stacking. Within this scheme, configurations 1-R60-DU, 4-R60-DU, and 2-R60-UD correspond to the experimentally observed H1 and L1 stackings in the high- (HT) and low-temperature (LT) phases, and their common stacking C2 of bulk $Nb_3X_8$, respectively (Fig. S3) [34,35,57–59].

Figures Fig. 2(f)-2(h) and S5 depict the total energies of all 24 $Nb_3X_8$ bilayer stackings in the NM, interlayer FM, and interlayer AFM configurations, while Fig. S6 presents the corresponding interlayer distance. The four halides exhibit halogen dependent trends in their relative energy profiles. For $Nb_3Cl_8$, $Nb_3Br_8$ and $Nb_3I_8$, the global energy minimum corresponds to the 2-R60-UD stacking (C2), whereas $Nb_3F_8$ favors the 3-R60-UD (AS) configuration over C2. The preference of C2 in $Nb_3Cl_8$ to $Nb_3I_8$ rationalizes the experimentally observed bilayer dimerization (with the C2 stacking) in its LT and HT bulk forms. By contrast, $Nb_3F_8$ is expected to adopt a different pairing registry. Because the C2 (or AS) stacking adopts a UD polarization alignment, the stacking registry between paired C2 bilayers in bulk $Nb_3X_8$ must be of the DU alignment. Among all DU configurations (yellow in Figs. 2f to 2h, and S5), the L1 stacking is consistently the most stable, while H1 is the second-lowest in energy only for $Nb_3Cl_8$. For DU stackings related to L1 by in-plane translation (i.e. without rotation), the most stable registry is 2-R60-DU for $Nb_3F_8$ and 3-R60-DU for $Nb_3Br_8$ and $Nb_3I_8$. These results suggest that bulk $Nb_3F_8$ in its LT phase adopts interleaved sequence of AS and L1 bilayers, a geometry distinct from the C2-based stacking sequences found in the heavier halides.

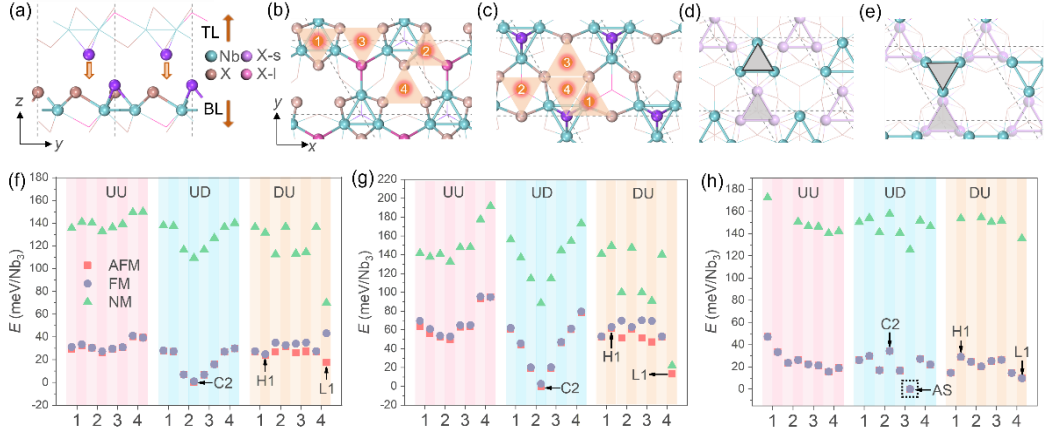

Fig. 2. (a) Side view of the fully relaxed atomic structure of a $Nb_3X_8$ bilayer. Interfacial X-s atoms of the top layer, and all interfacial X and Nb atoms of the bottom layer are highlighted, while other atoms are represented using lines. Brown arrows indicate the polarization directions of the top (TL) and bottom (BL) layers, respectively. (b-c) top-views of the atomic structure of the bottom layer in up- and down-polarizations, respectively. Shadowed numbers 1 to 4 mark four inequivalent X-hollow positions at the interface. All X atoms in the surface sublayer are blurred using line models. (d-e) Schematics of untwisted (R0, d) and 60° twisted (R60, e) bilayers. All X atoms are blurred using lines to highlight the relative position of top (light sea green) and bottom (light purple) layer Nb atoms. Gray shadowed triangles indicate $Nb_3$ trimers. (f-h) Energies of all 24 stacking configurations for bilayer $Nb_3Cl_8$ (f), $Nb_3I_8$ (g) and $Nb_3F_8$ (h) with a 1.2 eV $U$ value (h). Square, circle, and triangle symbols represent interlayer AFM, FM, and NM, respectively. Notation ZZ represents the intralayer zigzag magnetic configuration. Red, blue, and yellow shadowed blocks represent polarization directions of UU, UD and DU, respectively. The light and dark shaded blocks represent R0 and R60, respectively. The energy of the most stable stacking in each material is set as the reference (zero) energy.

In $Nb_3F_8$ bilayers, the most stable stacking is AS (3-R60-UD), rather than the C2 (2-R60-UD) configuration that dominates in $Nb_3Cl_8$, $Nb_3Br_8$, and $Nb_3I_8$. This contrast highlights the distinct structural preference of $Nb_3F_8$. In the UD configurations, the interfacial halogen atoms form four inequivalent triangles [highlighted by orange shadows in Fig. 2(c)], whose side lengths are listed in Table 2. In $Nb_3Cl_8$, $Nb_3Br_8$, and $Nb_3I_8$, triangle 2 exhibits the longest side length, resulting in the lowest electronic density. When the X-s atom in the top layer is positioned at the center of the triangle 2, the Coulomb repulsion between the X atoms within the triangles and the X-s atoms is minimized. This, in turn, results in the lowest energy for the 2-R60-UD stacking and the smallest interlayer distance [Fig. S6]. In contrast, in $Nb_3F_8$, the triangles undergo disproportionation, leading to the shortest side length for triangle 2 (2.82 Å) and the longest for triangle 3 (3.12 Å). Consistent with the relationship established earlier, the longest side length of triangle 3 gives rise to its lowest electron density, thereby resulting in the lowest energy and interlayer distance when the upper X-s atom is positioned on triangle 3. Consequently, in $Nb_3F_8$, the 3-R60-UD (AS) configuration exhibits the lowest energy.

We next take bilayer $Nb_3Cl_8$ in the representative L1 (4-R60-du) stacking configuration as a prototype to elucidate the role of interlayer coupling in modulating electronic structure. In previous studies, the L1 stacking configuration was considered to possess a NM ground state[32,60]; accordingly, we first examine the NM case. Figure 3(a) plots a spin non-polarized bandstructure of bilayer $Nb_3Cl_8$. Upon stacking two monolayers, the partially occupied flat bands originally present in isolated monolayers split into fully occupied valence band (VB1) and a completely empty conduction band (CB1). Both are spin-degenerate and located near the Fermi level. Wavefunction norm visualizations indicate that VB1 corresponds to a bonding state primarily localized on interfacial Cl atoms [ Figs. 3(b)], while CB1 is antibonding between two interfacial Cl sublayers [Figs. 3(c)]. This interlayer orbital hybridization corresponds to the energy splitting of flat bands in the nonmagnetic bilayer.

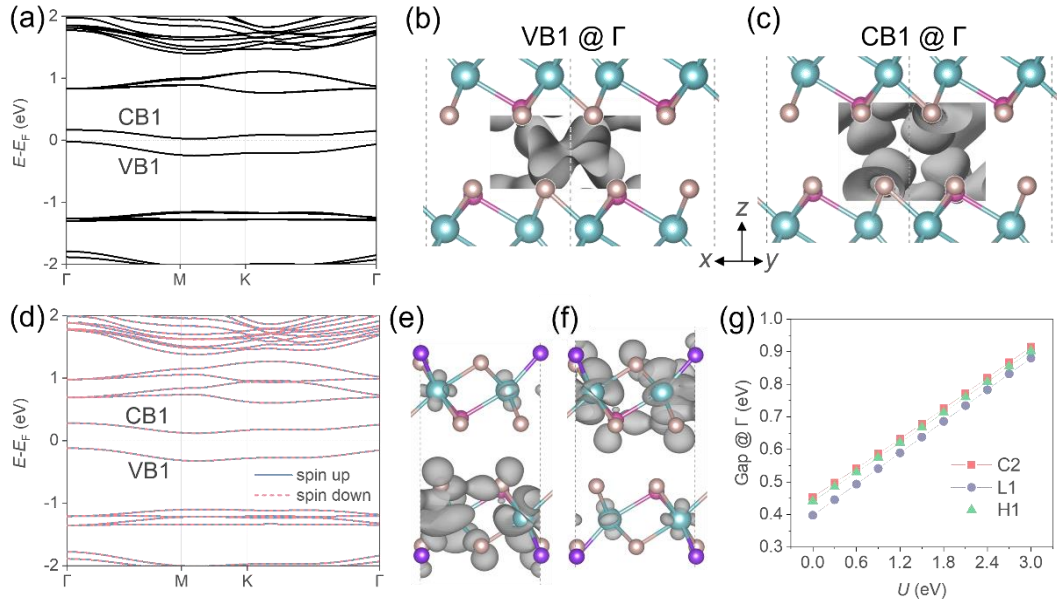

Fig. 3 (a) Electronic band structure of $Nb_3Cl_8$ with L1 stacking in the NM state. (b-c) Side views of $Nb_3Cl_8$ with L1 stacking. (d) Electronic band structure of $Nb_3Cl_8$ with L1 stacking in the interlayer AFM state. The isosurface represents the visualized wavefunction norms of the VB1 (b) and CB1 (c) at the $\Gamma$ point. The isosurface value was set to $3\times10^{-4}$ e/Bohr$^3$. (e-f) Visualized wavefunction norms of the VB1 for spin-up (g) and spin-down (h) components. The isosurface value was set to $5\times10^{-4}$ e/Bohr$^3$. (g) Band gaps at the $\Gamma$ point as a function of Hubbard $U$, calculated using interlayer AFM configurations.

In comparison to Fig. 3(a), we plotted the corresponding spin polarized bandstructure with the interlayer AFM coupling in Fig. 3(d). This AFM state is 19.4 meV/$Nb_3$ more stable than its spin non-polarized NM state. Similar to the NM case, the electronic structure near the Fermi level comprises two sets of flat bands, namely a fully occupied VB1 and an empty CB1 band, separated by a bandgap of 0.40 eV at the $\Gamma$ point. Each band set consists of two energetically degenerate but spatially distinct spin-polarized states. Visualized wavefunction norms at $\Gamma$ clearly demonstrate a layer-dependent spin locking that the spin-up component of VB1 predominantly localizes on the bottom layer [Fig. 3(e)], whereas the spin-down component resides mainly on the top layer [Fig. 3(f)]. This layer-spin locking is a general feature for all bands near the Fermi level. Specifically, the spatial wavefunction distributions of spin-up and spin-down states in CB1 [Figs. S8(a,b)] closely mirror those of spin-down and spin-up states in VB1, respectively. Consequently, each layer hosts energetically separated spin-polarized states with opposite spin alignments between layers, consistent with the characteristic of AFM-coupled Mott insulator. Furthermore, the bandgap at Γ exhibits a linear dependence on the applied Hubbard $U$ value [Fig. 3(i)], clearly identifying it as a Mott–Hubbard gap, with VB1 and CB1 corresponding to the lower Hubbard band (LHB) and upper Hubbard band (UHB), respectively. Such a linear gap-$U$ dependence is universal across all 24 stacking configurations of the four $Nb_3X_8$ bilayers studied, as illustrated by representative stacking orders C2, L1, and H1 of bilayer $Nb_3Cl_8$ in Fig. 3(g). These results highlight the robustness of these Mott insulator characteristics in $Nb_3X_8$ bilayers.

We use the $Nb_3Cl_8$ L1 and C2 stacking configurations as examples to investigate the impact of different polarization alignments on the electronic structure. Fig. 4(a-b) depict the structure and interlayer differential charge density (DCD) for the L1 stacking. At the interface, the two triangulars labeled as triangle 2 [Figs. 2(b), light brown atoms in the dashed box in Fig. 4(b)] from the top and bottom layers are aligned with a relative 60° twist. Within these triangles the electronic density is depleted, with the excess charge redistributed mainly to regions outside the triangles and partially into the interlayer region. The reduced charge density inside the triangles lowers the Coulomb repulsion among the three X atoms and hence shortens the triangle side length (from 3.42 Å in the monolayer to 3.34 Å in the L1 stacking). The interlayer charge accumulation enhances the electrostatic attraction between the two layers. In the C2 stacking configuration [Figs. 4(c-d)], the X-s atoms exhibit charge depletion, while the three nearest X atoms (assigned to triangle 2 in Fig. 2(c), light brown atoms in dashed boxes in Fig. 4(d)) in the opposing layer undergo charge accumulation. This charge redistribution gives rise to the formation of local dipoles oriented along the vertical direction. These oppositely aligned dipoles enhance the interlayer attraction via electrostatic interactions.

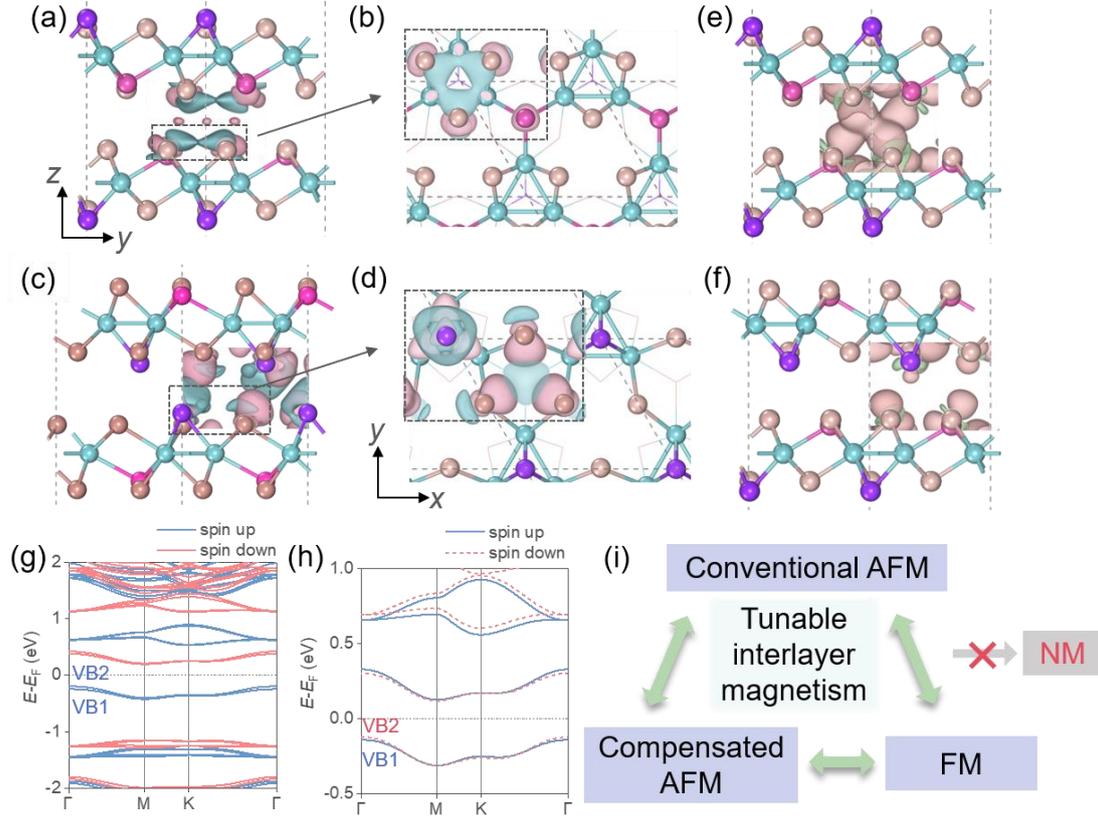

Fig. 4 (a) Side view of the L1 stacking of $Nb_3Cl_8$. The isosurface represents the interfacial differential charge density (DCD), with red indicating charge accumulation and blue indicating charge reduction. (b) Top view of the DCD from the dashed box in (a). (c-d) Same as (a-b) but for the C2 stacking of $Nb_3Cl_8$. (e-f) Side view of the L1 (e) and 4-R60-UU (f) stacking of $Nb_3Cl_8$. The isosurface represents the spin density in the interlayer FM coupling configuration, with orange for spin-up and green for spin-down. (g) Band structure of $Nb_3Cl_8$ with 4-R60-UU stacking in the FM configuration. (h) Band structure of $Nb_3Cl_8$ with AA (1-R0-UU) stacking in the AFM configuration. (i) Illustration of tunable interlayer magnetism in the $Nb_3X_8$ system. Interlayer DCDs are shown with an isosurface value of $1\times10^{-4}$ e/Bohr$^3$. Spin densities are shown with an isosurface value of $5\times10^{-4}$ e/Bohr$^3$.

We further examine how interlayer coupling modulates the favored magnetic configuration in $Nb_3X_8$ bilayers. Total energy comparisons among different magnetic configurations [Fig. 2 (f-h) and Fig. S5] reveal that most (71 out of 96) stacking configurations strongly favor interlayer AFM, while 22 out of 96 exhibit nearly degenerate interlayer AFM and FM states, with the energy difference between them being less than 0.1 meV/$Nb_3$. However, interlayer FM becomes more stable in the 4-R60-UU stacking in $Nb_3X_8$ (X = Cl, Br, I) [Fig. 4(f)]. To illustrate the mechanism of interlayer magnetic coupling, we use the L1 and 4-R60-UU stackings of $Nb_3Cl_8$ as prototypical examples. In $Nb_3Cl_8$, the L1 stacking configuration exhibits the largest energy difference $E$(FM) − $E$(AFM) = 23.4 meV/Nb. The spin density of the FM configuration in the L1 stacking [Fig. 4(e)] exhibits significant interlayer overlap, which enhances Pauli repulsion. This Pauli repulsion favors interlayer AFM coupling, which outweighs the FM tendency induced by interlayer hopping and thus stabilizes a relatively strong interlayer AFM ground state in the L1 stacking. [61]. In contrast, the interlayer overlapping is largely reduced in the 4-R60-UU configuration [Fig. 4(f)], leading to weaker Pauli repulsion. As a result, interlayer hopping dominates over Pauli repulsion the interlayer magnetic interaction, giving rise to interlayer FM coupling[61]. We plotted its FM band structure in Fig. 4(g). The splitting of the two VBs (noted VB1 and VB2) at the $\Gamma$ point is 49 meV (42 meV for the two CBs, noted CB1 and CB2), while VB1 and VB2 are degenerate at the K point. The visualized wavefunction at the K point shows that the spin-up component of VB1 and the spin-down component of CB1 are identical, as well as for VB2 and CB1 [Fig. S9(b-e)], which is consistent with the characteristics of a Mott insulator and confirms that the system retains its Mott insulator character. In summary, the interlayer magnetic coupling is determined by the competition between interlayer Pauli repulsion (which favors antiferromagnetism, AFM) and interlayer hopping (which favors ferromagnetism, FM). Since the orbitals contributing to the magnetic moment in $Nb_3X_8$ are predominantly distributed within the Nb atomic layers, the interlayer coupling is relatively weak. As a result, the wavefunctions of opposite spins from the two layers cannot overlap interlayer to form covalent bonds, leading to a NM state with relatively high energy.

Those 71 energetically favored AFM configurations can be categorized into, at least, two types of different spin groups. In every configuration with a R60 twisting angle in anti-parallel polarization alignments (up to eight configurations for each halide), an inversion symmetry connects the two spin sublattices, resulting in conventional AFM states with spin-degenerate bands developing throughout the Brillouin zone (BZ) [see Fig. S4(c) for the L1 stacked $Nb_3Cl_8$ bilayer]. However, every UU-polarized or every R0 (untwisted) configuration (totally up to 16 for each halide) exhibits an AFM configuration where no symmetry relates the two spin sublattices. The lack of symmetry yields a compensated AFM [61] state with appreciable spin splitting between the spin-up and -down bands throughout the entire Brillouin zone. The simplest structure of compensated AFM is the AA stacked configurations (1-R0-UU). Figure 4(l) shows the band structure of a compensated AFM AA stacking of $Nb_3Cl_8$, where the maximum splitting of up to 26 meV occurs among the four VBs and CBs near the Fermi level, particularly in the VB1 and VB2 at the Γ point. Therefore, $Nb_3X_8$ bilayers exhibit stacking tunable interlayer magnetism, ranging from FM, AFM-FM degeneracy, conventional AFM, to compensated AFM. Although purely electric-field-driven polarization switching in $Nb_3Cl_8$ remains challenging, combining electric fields with thermal activation, tensile strain, charge doping, or scanning-probe-tip manipulation offers feasible pathways for tuning interlayer magnetism.

Previous magnetic susceptibility measurements report that bulk $Nb_3Cl_8$ and $Nb_3Br_8$ behave as Curie-Weise paramagnets at high temperature but appear NM in their LT phases [32,34,35]. Two microscopic mechanisms were proposed to explain the origin of this NM behavior, namely inter-layer charge disproportionation[34] and formation of band insulator through interlayer interactions [35,63]. Although DFT is not designed to fully capture PM, it can directly describe non-magnetic states arising from either mechanism. Because both mechanisms require interlayer coupling, the smallest structural motif that can host them is a bilayer. Our exhaustive DFT calculations for all bilayers of four $Nb_3X_8$ compounds, however, reveal neither of these effects in any stacking considered. We further optimized the geometric structures of the LT phase of bulk $Nb_3Cl_8$ and $Nb_3Br_8$ using the NM configuration and compared their energies in NM, and interlayer AFM and FM configurations. In every case, the interlayer AFM state is energetically more favored than the NM and interlayer FM states. The energy difference between the NM and interlayer AFM states ($E_{NM} - E_{AFM}$) increases monotonically with the on-site Hubbard $U$, consistent with our results for bilayer calculations. These results indicate that, within the DFT(+U) framework, the ground state of LT-phase bulk $Nb_3Cl_8$ and $Nb_3Br_8$ is most plausibly an interlayer AFM Mott insulator. This assignment is not inconsistent with the nearly featureless LT susceptibility reported experimentally. Should the true ground state prove to be strictly non-magnetic, its origin must involve physics beyond standard DFT (+U), which calls for further investigations.

In summary, we found that in Nb$_3$X$_8$ bilayers, all magnetic ground states are interlayer AFM, with a few degenerate FM states, exhibiting tunable interlayer AFM. However, whether in bilayers or bulk, the NM states have much higher energies than the magnetic ground states, suggesting that previous understanding may need revision or that beyond-DFT-level theoretical calculations are required for a better understanding. Through the analysis of the electronic structure and the linear dependence of band splitting with $U$, we found that all considered bilayers are Mott insulators, indicating robust Mott insulating behavior in this system. Regardless of the interlayer magnetic coupling, the Mott insulating nature persists, which contrasts with the previous assumption that the system could exhibit a NM state. The robustness of the Mott insulator state and the absence of an NM phase in Nb$_3$X$_8$ contrast with NbSe$_2$. In NbSe$_2$, the interfacial Se atoms have out-of-plane $p_z$ orbitals, which results in strong interlayer coupling, allowing unpaired electrons to form interlayer bonds[63]. In contrast, in Nb$_3$X$_8$, the states near the Fermi level are primarily derived from Nb orbitals, with a much smaller contribution from halogen atoms. As a result, the interlayer coupling in Nb$_3$X$_8$ is much weaker. Our study provides unexpected results and insights into the role of interlayer coupling in modulating correlated insulators, contributing to the further development of this field.